\documentclass[conference]{IEEEtran}
\IEEEoverridecommandlockouts
\usepackage{cite}
\usepackage{amsmath,amssymb,amsfonts}
\usepackage{algorithm,algorithmic}
\usepackage{graphicx}
\usepackage{textcomp}
\usepackage{booktabs}
\usepackage{braket}
\usepackage{tabularx}
\def\BibTeX{{\rm B\kern-.05em{\sc i\kern-.025em b}\kern-.08em
    T\kern-.1667em\lower.7ex\hbox{E}\kern-.125emX}}

\usepackage{titlesec}
\titlespacing*{\section}{0pt}{2pt}{2pt}
\titlespacing{\subsection}{0pt}{2pt}{2pt}

\usepackage{lineno,hyperref}
\hypersetup{
    colorlinks   = true,
    linkcolor    = blue,
    citecolor    = red
}
\usepackage{bm}
\usepackage{mathrsfs}
\usepackage{graphicx}
\usepackage{subfigure}
\usepackage[dvipsnames]{xcolor}
\usepackage{nicefrac}
\usepackage{array}
\usepackage{float}
\usepackage{balance}
\usepackage{multirow} 
\usepackage{quantikz}
\usepackage{tikz}
\usetikzlibrary{arrows.meta,shapes,positioning,fit,backgrounds,calc}
\definecolor{borderpurple}{RGB}{128, 55, 155}
\definecolor{lightpurple}{RGB}{233, 225, 242}
\definecolor{medpurple}{RGB}{201, 176, 242}

\usepackage{adjustbox}
\usepackage[flushleft]{threeparttable} 

\makeatletter
\newcommand*{\transpose}{%
  {\mathpalette\@transpose{}}%
}
\newcommand*{\@transpose}[2]{%
  \raisebox{\depth}{$\m@th#1\intercal$}%
}
\newcommand{\thickhline}{%
    \noalign {\ifnum 0=`}\fi \hrule height 1pt
    \futurelet \reserved@a \@xhline
}
\newcolumntype{"}{@{\hskip\tabcolsep\vrule width 1pt\hskip\tabcolsep}}

\makeatother

\ifCLASSINFOpdf

\else

\fi

\begin{document}
\renewcommand{\ttdefault}{cmtt}
\bstctlcite{IEEEexample:BSTcontrol}


\title{Transient-Stability-Aware Frequency Provision in IBR-Rich Grids via Information Gap Decision Theory and Deep Learning}

\author{\IEEEauthorblockN{Amin Masoumi and Mert Korkali}
\IEEEauthorblockA{\textit{Department of Electrical Engineering and Computer Science}\\ 
\textit{University of Missouri} \\
Columbia, MO 65211 USA \\
e-mail: \{\texttt{am4n5,korkalim\}@missouri.edu}}}

\maketitle

\begin{abstract}
This paper introduces a framework to address the critical loss of transient stability caused by reduced inertia in grids with high inverter-based resource (IBR) penetration. The proposed method integrates a predictive deep learning (DL) model with information gap decision theory (IGDT) to create a risk-averse dispatch strategy. By reformulating the conventional virtual inertia scheduling (VIS) problem, the framework uses early predictions of post-fault dynamics to proactively redispatch resources, ensuring the system's center of inertia remains stable under worst-case contingencies. Validated on the IEEE 39-bus system with $70\%$ IBR penetration, the proposed approach prevents system collapse where a conventional VIS strategy fails, ensuring frequency stability at a cost increase of only $5\%$.
\end{abstract}


\section{Introduction}\label{sec:sec1}
Power systems are undergoing a significant evolution to incorporate inverter-based resources (IBRs), such as wind turbines, solar photovoltaics, and battery storage. This change is driven by environmental concerns, technological advancements, and economic benefits \cite{bevrani2017microgrid}. A crucial driver is the economic benefit of IBRs, which possess a lower levelized electricity cost than traditional synchronous generators (SGs). Two primary control paradigms for the integration of IBRs into power systems include grid-following (GFL) and grid-forming (GFM) modes. GFL IBRs are synchronized with the grid and function as current sources for power injection. In contrast, GFM IBRs are notable for maintaining stable voltage and frequency without the use of SGs \cite{karimjee2025analysis}. The replacement of SG with IBR extensively diminishes natural inertia, adversely affecting the system's frequency and transient stability \cite{wang2025dynamics}. To mitigate this issue, methods have been devised to emulate the inertia and damping properties of synchronous generators (SGs). This has resulted in the development of virtual synchronous generators (VSGs) and virtual inertia scheduling (VIS) \cite{hu2022grid}. According to \cite{yang2022distributed}, optimal parameter configurations are contingent on system operating conditions, disturbance characteristics, and stability requirements established by balancing authorities (BAs). A major issue stems from the swift release of synthetic kinetic energy during frequency incidents, which can exhaust the headroom reserved for frequency stabilization. This complication is especially pronounced when the units operate near full capacity, which offers minimal opportunity for additional power inputs during interruptions \cite{milano2018foundations}. Reduced inertia in IBR-rich grids accelerates frequency dynamics and shortens the critical clearing time (CCT), increasing the susceptibility of the grid to transient instability \cite{tielens2016relevance} and subsequent cascading failures. In particular, flexible alternating current transmission system (FACTS) devices \cite{singh2022congestion}, such as static synchronous compensators (STATCOMs), series capacitors, and high-voltage direct-current (HVDC) systems \cite{yan2023frequency}, are current methods to boost system stability by hardening the grid, optimizing power flow on congested lines, and ensuring rapid energy transfer. However, their high capital investment and intricate control mechanisms hinder broad adoption, and operating expenses over a long period may render them impractical for many utilities. 
In contrast, aggregated electric vehicle (EV) fleets offer substantial potential for secondary frequency regulation due to lower investment costs and distributed deployment \cite{bakhuis2025exploring}. However, a significant hurdle is the conflict between locational marginal pricing by the transmission system operator and the incentives for EV owners. This mismatch hinders optimal EV grid support unless mandated by BAs. The conflict stems from the balance between energy consumption and battery degradation. BA intervention, therefore, can reduce social welfare and public acceptance of EV grid services \cite{shao2015hierarchical}, challenging EV-based frequency and transient stability management in environments with high IBR penetration. Imagine a scenario with a significant load increase (e.g., $60\%$) at one of the intersections, followed by faults on buses susceptible to three-phase short circuits. The CCT is then affected by economic dispatch signals in a low-inertia, IBR-rich grid. A slight reduction in CCT could trigger three-phase short circuits, resulting in cascading failures and potentially leading to system collapse. Allowing only $1$-$2\%$ load factor participation and load shedding is insufficient due to the impact on the frequency nadir, which could lead to underfrequency load shedding (UFLS). Although rapidly discharged electric vehicle fleets can temporarily address specific issues, economic and social factors pose significant challenges. A significant challenge in grids with high IBR penetration is predicting load loss following a transient instability event. The application of advanced algorithms, such as deep learning (DL), to forecast this load loss can significantly improve the coordination of automatic generation control (AGC) signals, thus mitigating the impact of the frequency nadir and creating opportunities to restore the load. In the absence of such predictive information, economic dispatch (ED) operators and AGC systems often react too aggressively to these disturbances. This study addresses this issue by reformulating the cost function to incorporate the risk of system collapse directly into real-time economic dispatch (RTED), which also informs a long-term strategy for coordinating EV fleets. To achieve this, we introduce a robust optimization framework that leverages the predictive power of DL \cite{masoumi2025deep} and the uncertainty management capabilities of information gap decision theory (IGDT) \cite{ben2006info}. The DL-IGDT method establishes a risk-averse strategy by bounding the cost function against worst-case scenarios, producing superior solutions compared to unbounded functions that can lead to aggressive and penalty-driven actions. To our knowledge, this work is the first to integrate DL along with IGDT into the RTED process. Our primary contributions are as follows:
\begin{itemize}
    \item Developing a comprehensive framework for addressing transient stability awareness in IBR-rich grids;
    \item Designing detailed robust optimization by combining DL (early prediction) and IGDT; and
    \item Designing novel risk-aware technique for system operators to assess the economic value in transition toward IBR-dominated grids.
\end{itemize}

The remainder of this paper is organized as follows: Section \ref{sec:sec2} constructs the conventional VIS optimization problem, detailing the cost functions for various generation resources and their associated operational constraints. Section \ref{sec:sec3} introduces the proposed framework for assessing transient stability, which reformulates the VIS problem using IGDT and employs a DL model to predict and quantify risk induced by contingencies. Section \ref{sec:sec4} presents the results of validating the proposed method on the IEEE 39-bus test system. Finally, Section \ref{sec:sec5} concludes the paper.

\section{VIS and the Cost Function Description}\label{sec:sec2}
In this section, the optimal VIS problem of the IBR-rich grid is constructed. The total cost function is given by \cite{she2023virtual}
\begin{subequations}
\begin{align}
&\text{Min } \text{Cost}^{\text{total}} = \text{Cost}^{P_g} + \text{Cost}^{\text{IBR}} +\text{Cost}^{\text{EV}} \label{eq:eq1}\\ 
&\text{Cost}^{P_g} =\!\!\! \sum_{g\in\Omega_G} (c_2 P_g^2 + c_1 P_g + c_0  + c_g^U R_g^U  + c_g^D R_g^D) \label{eq:eq2}\\
&\text{Cost}^{\text{IBR}} =\!\!\! \sum_{{\text{IBR}}\in\Omega_I} (c_5 P_{\text{IBR}}^2 + c_4 P_{\text{IBR}} + c_3  + c_6^M M_{\text{IBR}} + c_7^D D_{\text{IBR}}) \label{eq:eq3}\\
&\text{Cost}^{\text{EV}} = \!\!\! \sum_{\text{EV}\in\Omega_E} \lambda^U P_{\text{EV}}^U  + \lambda^D P_{\text{EV}}^D \label{eq:eq4}
\end{align}
\end{subequations}
\noindent where (\ref{eq:eq2})--(\ref{eq:eq4}) are the cost functions of conventional SGs, IBRs, and EV fleets. ${P_g}$, $R_g^U$, and $R_g^D$ are scheduled power, regulation-up reserve, and regulation-down reserve of the SG (decision variables), respectively. In addition, $P_{\text{IBR}}$, $M_{\text{IBR}}$, and $D_{\text{IBR}}$ are scheduled power, scheduled virtual inertia, and scheduled damping of IBRs, respectively. Furthermore, $P_{\text{EV}}^U$ and $P_{\text{EV}}^D$ represent the scheduled regulation-up reserve and regulation-down reserve of the EV fleets, respectively. In addition, $c_1, \dots, c_7$, $\lambda^U$, and $\lambda^D$ are constant coefficients of the IBR-rich grid, regulation-up reserve, and regulation-down reserve prices of EV fleets. Furthermore, the constraints are modeled as
\begin{subequations}
\begin{align}
&\Delta D^U = \sum_{g\in\Omega_G} R_g^U  + P_{\text{EV}}^U \label{eq:eq5}\\
&\Delta D^D = \sum_{g\in\Omega_G} R_g^D  + P_{\text{EV}}^D \label{eq:eq6}\\
&\sum_{g\in\Omega_G} P_g + \sum_{{\text{IBR}}\in\Omega_I} P_{\text{IBR}} = P_d \label{eq:eq7}\\
&\sum_{{\text{IBR}}\in\Omega_I} M_{\text{IBR}} = M_{\text{req}} \label{eq:eq8}\\
&\sum_{{\text{IBR}}\in\Omega_I} D_{\text{IBR}} = D_{\text{req}} \label{eq:eq9}\\
&R_g^U + P_g \leq P_g^{\text{max}}, \ \forall g \in \Omega_G \label{eq:eq10}\\
&P_g^{\text{min}} \geq  P_g - R_g^D , \ \forall g \in \Omega_G \label{eq:eq11}\\
&P_{g,t} - P_{g,t-1} \leq  \gamma_g^U, \ \forall g \in \Omega_G \label{eq:eq12}\\
&P_{g,t-1} -P_{g,t}  \leq  \gamma_g^D, \ \forall g \in \Omega_G \label{eq:eq13}\\
\begin{split}
{P}_{l}^\text{min} \leq \sum_{g\in\Omega_{G}} \text{GSF}_{g,k} \left(P^{\text{sch}}_{g}-D_{g}\right) &\leq {P}_{l}^\text{max},\\
\forall k \in \Omega_{K}, \ \forall g \in \Omega_{G}
\end{split}
\label{eq:eq14}\\
&P_{\text{EV}}^U  \leq P_{\text{EV}}^{U,\text{max}}, \ \forall \text{EV} \in \Omega_E \label{eq:eq15}\\
&P_{\text{EV}}^D  \leq P_{\text{EV}}^{D,\text{max}}, \ \forall \text{EV} \in \Omega_E \label{eq:eq16}\\
&M_{\text{IBR}} \leq {M}_{\text{IBR}}^{\text{max}} \label{eq:eq17}\\
&D_{\text{IBR}} \leq {D}_{\text{IBR}}^{\text{max}} \label{eq:eq18}
\end{align}
\end{subequations}
\noindent where (\ref{eq:eq5}) and (\ref{eq:eq6}) demonstrate zonal up and down imbalances compensated by regulation-up and down reserve values of SGs and EV fleets. Constraints (\ref{eq:eq7})--(\ref{eq:eq9}) are used to formulate the power balance, virtual inertial balance (zonal inertia), and virtual damping balance (zonal damping), respectively. Constraints (\ref{eq:eq10}) and (\ref{eq:eq11}) indicate cleared scheduled and reserve transactions in the RTED-VIS
routine. Constraints (\ref{eq:eq12}) and (\ref{eq:eq13}) represent $10$-minute ramp limits for SGs. Equation (\ref{eq:eq14}) illustrates the line flow restriction, with $\text{GSF}$ denoting the generation shift factor. Constraints (\ref{eq:eq15}) and (\ref{eq:eq16}) impose the maximum and minimum output limits of the EV fleet reserve. Constraints (\ref{eq:eq17}) and (\ref{eq:eq18}) express the maximum limitations of virtual inertia and virtual damping, respectively.

\section{Proposed Transient-Stability-Aware VIS via DL-IGDT}\label{sec:sec3}
In this section, we will establish a formulation for the VIS problem that is informed by considerations of transient stability. Initially, we will discuss the challenges associated with the current formulation and its implications for the CCT. Eventually, a detailed exposition of the proposed reconceptualization of the transient-stability-aware framework will be provided.

\subsection{Problem Description}
A power system's resilience to major disturbances, such as three-phase short-circuit faults, can be quantified by its CCT---the maximum fault duration the system can endure without losing synchronism. During a fault, the deviations in the generator rotor angles increase, and if the fault is not cleared within the CCT, the system can suffer an out-of-step (OOS) condition, which can lead to a collapse.

The integration of IBRs compromises stability in two compounding ways. First, by displacing conventional synchronous generators, IBRs reduce the system's natural inertia, which shortens the CCT. Second, unlike synchronous generators that supply high fault currents ($5$--$10\times$ their rating) to enable rapid fault isolation, IBRs limit their current output (to around $1.2$--$2\times$ their rating) to protect their power electronics. This current-limiting behavior can prolong fault clearing times, especially in systems with legacy protection schemes.

This creates a dangerous dynamic in IBR-rich grids: the time available to clear a fault (CCT) decreases, while the actual time it takes to clear it may increase, significantly elevating the risk of instability. As shown in Fig. \ref{fig:Fig1} (unstable scenario), this can lead to a loss of synchronism during severe disturbances. While post-fault control methods exist, their effectiveness is limited by computational demands. A more robust solution is a predictive approach that can proactively adjust grid resources before an OOS condition develops. To this end, we develop a risk-averse cost function that uses early prediction to stabilize grids with high IBR penetration.

\subsection{Transient-Stability-Aware Formulation of VIS }
In this study, we investigate an early prediction strategy by continuously monitoring the status of grids with high penetration of IBRs. Our objective is to reassess the VIS problem. By predicting the nature of contingencies---identifying whether they result in stable or unstable faults---we evaluate their potential effects. We particularly focus on the frequency deviation, represented as $\Delta f$, in the post-fault trajectory, which forms the basis of our proposed predictive algorithm. This technique is predicated on estimating the amount of load disruption by computing the area control error (ACE), utilizing the observed frequency deviation as
\begin{subequations}
\begin{align}
    \text{ACE} &= \Delta P_{\text{tie}} - 10B \cdot \Delta f, \label{eq:eq19}\\
    \hat {P}_d &= P_d + \text{ACE} \label{eq:eq20},
\end{align}
\end{subequations}
\noindent where $\Delta P_{\text{tie}}$, $B$, $\Delta f$, and $ \hat {P}_d$ are the power interchange of the tie lines, frequency bias, frequency deviation from the actual frequency, and uncertain interrupted load, respectively. Note that $B$ is a negative number. Also, $\hat {P}_d$ is equal to $ {P}_d$ in the normal operation mode, that is, the typical VIS problem. In other words, the impacted load is the uncertain variable of the VIS problem under a transient instability condition. This introduces IGDT as a remedy for uncertainty quantification. Based on its definition, IGDT aims to close the gap between the predicted value and possible deviations from the real-time value. IGDT quantifies the uncertainty as a fractional deviation from the predicted (interrupted) load as 
\begin{equation}
\label{eq:eq21}
\Im (\sigma , \hat{P}_{d}) = \left\{ P_{d} : \frac{|P_{d} - \hat P_{d}|}{\hat P_{d}} \leq \sigma \right\} , \ \sigma  \ge 0,
\end{equation}

\noindent which limits the gap ($\sigma$) to be proportional to the predicted value. We now formulate a robustness function ($R(C_r)$) to characterize the worst-case uncertainty bounds that an IBR-rich grid can withstand under contingency. In other words, the robustness function is the risk-averse strategy that prepares FP for extreme power system contingencies. Hence, the IBR-rich grid cost function is developed to include the robustness factor. Specifically, the proposed cost function for the VIS problem is the maximum optimal cost that the IBR-rich grid is willing to pay to avoid complete blackouts due to the OOS condition, which is given by
\begin{equation}
\label{eq:eq22}
R(C_r)=\max \left\{ \sigma : \left( \max_{P \in \Im(\sigma, \hat{P}_{d})} \text{Cost}^{\text{total}} \leq (1 + \theta)C_c \right) \right\}.
\end{equation}

Hence, the proposed cost function for the VIS problem is given by
\begin{equation}
\label{eq:eq24}
\begin{array}{l}
\max \left\{ {{\rm{Cos}}{{\rm{t}}^{{\rm{total}}}}} \right\} \le (1 + \theta ){C_c}\\
\text{subject to}\\
(2)-(4)
\end{array}
\end{equation}
\begin{equation}
\label{eq:eq25}
 P_{d} \leq (1+\sigma)\hat  P_{d}
\end{equation}

\noindent where (\ref{eq:eq22})--(\ref{eq:eq25}) constitute the proposed transient-stability-aware formulation of VIS. In addition, $\theta$ and $C_c$ are the probability of system collapse due to transient instability and the critical cost of the system before the operation of UFLS. Hence, the only parameters for initiating the optimization phase are the determination of $\hat P_d$ and $\theta$.

\subsection{Early Prediction of the Uncertainty and Risk Factor Quantification }
To predict $\hat P_d$, we apply a DL technique that benefits from the attention mechanism. In this regard, time-domain simulation (TDS), that is, Algorithm~\ref{alg1}, is designed to predict the impact of contingency by predicting the stability status and frequency deviation of the first swing, that is, the short-circuit fault in the three-phase scenario $\upsilon$ ($\xi^{\text{fault}, x^\upsilon}$) using TDS. Hence, the status of the transmission lines ($\xi^{\ell, \upsilon}$) is set to $0$ during a fault. Faults are simulated on all transmission lines, varying the duration of the fault ($\tau^\upsilon$), the location of the fault ($x^\upsilon$), and the system load level ($k^\upsilon$). They are selected in the ranges of $\tau^\upsilon=[0.06, 0.4]$ s, $x^\upsilon=[0, 100] \%$ and $k^\upsilon=[1,1.6]$ \rm{pu}. The faults are applied at $t_\text{start}=2$ s and their impact is monitored for durations of $5$ s and $7$ s, which are compatible with standard environments. For this study, we select the IEEE 39-bus (10-machine New England) system as the test case due to its capability to capture the oscillation in the rotor angles of bulk power systems. In addition, it includes critical lines and areas of concentrated power flow, which are ideal for studying cascading failures and assessing the effectiveness of preventive measures. Hence, TDS is performed after setting the inputs, including the duration of the fault, the location of the fault, and the system load, to assess the impact of rare event scenarios. Note that $\xi^{\ell, \upsilon}$ and $\xi^{\text{fault}, x^\upsilon}$ are the status of line $\ell$ and the fault status based on its location ($x^\upsilon$) on Line $\ell$, respectively.

\begin{algorithm}
 \caption{Modeling Three-Phase Short-Circuit Scenarios in a TDS Environment}
 \label{alg1}
 \begin{algorithmic}[1]
 \renewcommand{\algorithmicrequire}{\textbf{Input:}}
 \renewcommand{\algorithmicensure}{\textbf{Output:}}
 \REQUIRE $\tau^\upsilon; k^\upsilon; x^\upsilon$; $\xi^{\ell, \upsilon}, \ \ell=\{1,\dots,46\}; \xi^{\text{fault}, x^\upsilon}$
 \ENSURE TIS
 \\ \textit{Initialization}: $t_{\text{start}}^{\ell,\upsilon} = t_\text{start}, t_{\text{end}}^{\ell,\upsilon} = t_\text{start} + \tau^\upsilon, S_{{\text{rms}}}^\upsilon = k^\upsilon S_{{\text{rms}}}, \rm{timevector}=0:7$ s
 \FOR {$\upsilon = 1:V$}
    \STATE $S_{{\text{rms}}}^\upsilon = k^\upsilon S_{{\text{rms}}}$
    \FOR {$t \in \rm{timevector}$}
        \IF {($t \geq t_{\text{start}}^{\ell,\upsilon}$ \textbf{and} $t \leq t_{\text{end}}^{\ell,\upsilon}$)}
            \STATE $\xi^{\ell, \upsilon} = 0, \xi^{\text{fault}, x^\upsilon} = 1$
        \ELSE
            \STATE $\xi^{\ell, \upsilon} = 1, \xi^{\text{fault}, x^\upsilon} = 0$
        \ENDIF
    \ENDFOR
    \RETURN TIS
 \ENDFOR
 \end{algorithmic} 
\end{algorithm}

\begin{figure}[h!]
\centering
\includegraphics[width=.87\columnwidth,keepaspectratio=true,clip=true]{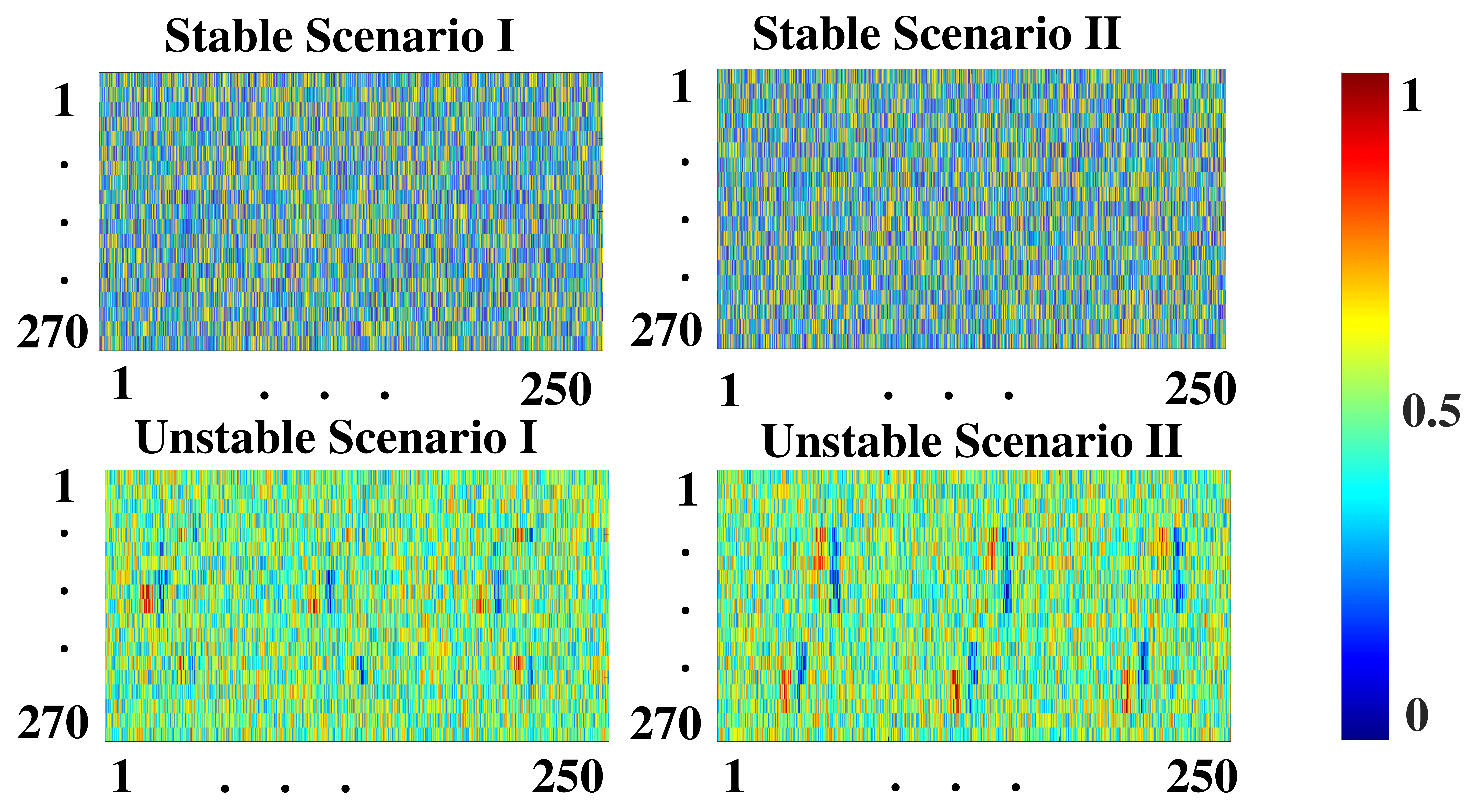}
 \vspace{-0.4cm}
\caption{The blurred colormap representation of the tabular data.}
\vspace{-0.2cm}
\label{fig:Fig1}
\end{figure}

For offline training, the proposed DL algorithm emphasizes rare event scenarios in TDS. Measurements are assumed to be collected on each generator bus, with a sampling frequency of $8$ ms ($\approx \frac{1}{{120}}$ s). The offline training process uses pre-fault and fault datasets to classify the OOS condition. This classification is crucial for online monitoring and evaluating the efficacy of identifying OOS conditions. A sliding window, with a length of $0.5$ s, is utilized to extract features for the classification of instability. This window captures pre-fault data ranging from $0.44$ s to $0.1$ s before fault clearance. The study examines a set of $27$ features, which are categorized into three sets: $9$ primary features, including instantaneous measurements (i.e., $i_d$, $i_q$, $v_d$, $v_q$, $\delta$, $\omega_m$, $T_e$, $P_g$, and $Q_g$), $9$ secondary, and $9$ tertiary features. The second set of features is derived by calculating the differences between consecutive samples in the primary set. The final set of features is determined on the basis of the differences in the rotor angles of the generators expressed as $\delta_{i,j} = |\delta_{i}-\delta_{j}|$. The \emph{transient instability status} (TIS) is assessed following the analysis of the effects of rare event scenarios on rotor angles throughout the $7$-s TDS with
\begin{equation}
\label{eq:eq26}
{\text{TIS = }}\left\{ \begin{gathered}
  \text{class 1}{\text{, if}} \ \frac{{360 - {\Lambda ^{\max }}}}{{360 + {\Lambda ^{\max }}}} < 0 \hfill \\
  {\text{class 0, if}} \ \frac{{360 - {\Lambda ^{\max }}}}{{360 + {\Lambda ^{\max }}}} > 0 \hfill \\
\end{gathered}  \right.
\end{equation}

\noindent where ${\Lambda ^{\max }} = \max \{|{\delta_i} - {\delta_j}|\},  \ i, j \in \{1,\dots,10\}$, is the maximum difference between angles of any two generators. We conducted TDS on the IEEE 39-bus system, employing a set of $27$ features for both pre-fault and fault conditions. Data were sampled every $8$ ms for a period of $0.5$ s, yielding $62$ time samples. Subsequently, these data were upsampled to a resolution of $250$ samples, resulting in a tabular data structure of dimensions $270 \times 250$. To visualize the differences in system behavior before and after contingencies, we utilized colormap intensity representations, as detailed in ~\cite{R2}. The intensity calculation of the blurred colormap over the tabular dataset is illustrated in Fig.~\ref{fig:Fig1}. We also identified two pairs of stable and unstable contingencies using the TIS. Notably, Fig.~\ref{fig:Fig1} demonstrates that unstable contingency pairs can be effectively differentiated from stable ones through colormap intensity, specifically in the ranges of $0$-$0.5$ and $0.5$-$1$ for stable conditions. Unstable scenarios are characterized by pronounced vertical density areas, which highlight the spatiotemporal effects of dynamic contingencies on the oscillatory behavior in power systems. 

\vspace{-0.3cm}

Certain features exhibit abnormal behavior during a fault, specifically in the blue and red areas. This behavior is primarily influenced by the dynamics of the IEEE 39-bus system and the sensitivity of these features to faults, resulting in spatial entanglement. The recurrence of this spatial entanglement in TDS indicates temporal dependence during fault clearance. 

For the current classification problem, we have chosen a convolutional neural network (CNN) as the optimal framework. The core challenge involves deciphering the spatiotemporal dynamics by linking the intrinsic behaviors that are common across various scenarios. We posit that the vertical and narrow anomalies, depicted as blurred colormap intensities across $270$ features, are consistently evident in two distinct unstable scenarios. Therefore, enhancing the intensity in these regions is likely to improve classification accuracy due to their shared visual characteristics. To address this, we propose an algorithm that integrates a CNN with an attention mechanism~\cite{attention} (denoted CNN-Att) and applies multiple filters to the tabular data. This methodology enables us to uncover the interdependencies of the spatial deep features common among the generators. The resulting volume data, consisting of tabular data concatenated with $5$ filters, are then fed into the CNN to extract the significance of the spatial features. Another critical challenge during contingencies is modeling the temporal interdependencies of the latent features. We leverage the attention mechanism to assess the correlation between pre-fault and fault time samples. This processed information is then passed into a dense feedforward neural network coupled with a softmax operator for the classification of TIS. The final step involves training a dense NN to predict the frequency of the first swing, using the estimated frequency to derive $\hat P_d$ according to (\ref{eq:eq20}).
CCT serves as the essential metric to assess the likelihood of transient instability ($\theta$) under specified operating conditions such as RTED-VIS, particularly in the context of contingency events. Within our probabilistic framework, we derive the probability of transient instability by examining the probability density function of fault durations that exceed the CCT threshold. This approach translates to a weighted sum of the probabilities associated with unstable scenarios characterized by fault durations that surpass the CCT. Consequently, the critical parameters for our probabilistic analysis include the probabilities of fault occurrences and their respective durations. In doing so, the probability of fault occurrence, i.e.,
\begin{equation}
    P_\ell(L) = \frac{{{\eta ^\ell}}}{\sum\limits_{q = 1}^\ell {{\eta ^q}} }
\end{equation}
is a function of the annual line fault rate ($\eta ^\ell$). We also generate the probabilities of the duration of the fault using two normal distributions, that is, $P_T(\tau^\ell) = \frac{1}{{2\sigma \sqrt {2\pi } }}{e^{ - \frac{{{{(\tau^\ell - 3.5)}^2}}}{{2{\sigma ^2}}}}} + \frac{1}{{2\sigma \sqrt {2\pi } }}{e^{ - \frac{{{{(\tau^\ell - 4)}^2}}}{{2{\sigma ^2}}}}}$. Therefore, the probability of transient instability can be defined as
\begin{subequations}
\begin{align}
P^\ell(\theta) &= \left\{ {\begin{array}{*{20}{l}}
{0\qquad \qquad \qquad {\rm{     }}\forall \tau^\ell  \in \{ \tau^\ell :\tau^\ell  < {\rm{CCT}}\} }\\
{{P_T}(\tau^\ell ){P_\ell }(L)\quad \forall \tau^\ell  \in \{ \tau^\ell :\tau^\ell  > {\rm{CCT}}\} }
\end{array}} \right.\label{eq:eq27} \\
\theta  &= \sum\limits_\ell^L {{P^\ell}(\theta )} \label{eq:eq28}
\end{align}
\end{subequations}

After acquiring $\theta$ and $\hat P^d$, we insert them into the transient stability-aware formulation (\ref{eq:eq24}) and proceed with optimization. 

\section{Results and Discussions}\label{sec:sec4}

We present a case study on the IEEE 39-bus test system characterized by a $70\%$ IBR penetration. The cost associated with load shedding is established at $\$1{,}000$/MWh. Furthermore, the critical cost ($C_c$) related to the proposed transient-stability-aware VIS is defined as $\$460$. The CCT of the system, evaluated under the aforementioned $70\%$ penetration of IBR, is recorded at $0.128$ s. The contingency arises from a three-phase short-circuit fault occurring at $2$ s on Line 33, which is located near Bus 26 and involves $95\%$ of the line's coverage, with a fault probability of $0.63$. In addition, the frequency incident affecting the power grid under analysis is related to the SG (Bus 33, with a rating of $1174.8$ [MVar]), which experiences a trip at $3$ s. Data acquisition and TDS have been implemented using the ANDES library. The environment for this study has been programmed in Python 3 and operates on a Google Compute Engine backend. 
\begin{table}[h!]
\vspace{-.2cm}
\caption{Accuracy of Classification Task regarding TIS}
\vspace{-.4cm}
\label{Table1}
\begin{center}
\begin{tabular}{c|c}
\hline
\hline
\textbf{Classifier} & \textbf{Accuracy} [$\%$]  \\
\hline

BiLSTM  & $95.2\%$ \\

BiGRU  & $91.47\%$ \\

Proposed CNN-Att  & $\bf{99}\%$\\
\hline
\hline
\end{tabular}
\end{center}
\vspace{-.2cm}
\end{table}


Table~\ref{Table1} illustrates the accuracy of the DL-based approach created for assessing TIS. In addition, we examined this method's capability in determining the frequency deviation linked to the first swing equation, resulting in a root-mean-square error (RMSE) of $0.019$. This is notably lower compared to other machine learning methods evaluated, such as support vector machine (SVM), Gaussian process regression (GPR), and decision trees (DT), which exhibited RMSE values of $0.073$, $0.254$, and $0.416$, respectively.
The regression component involves $3$ hidden layers with $400$, $300$, $300$ neurons, respectively. We have limited the accuracy of the results to estimate the frequency deviation based on the threshold to classify transient stability, specifically set at $90\%$. As evidenced by the results, the proposed CNN-Att algorithm outperforms all other DL counterparts. In addition, the center of inertia (COI) provides a detailed justification for evaluating the performance of the proposed methodology. 
\vspace{-0.1cm}
\begin{figure}[ht!]
\centering
\includegraphics[width=0.45\textwidth,trim=70px 70px 70px 70px,clip=true]{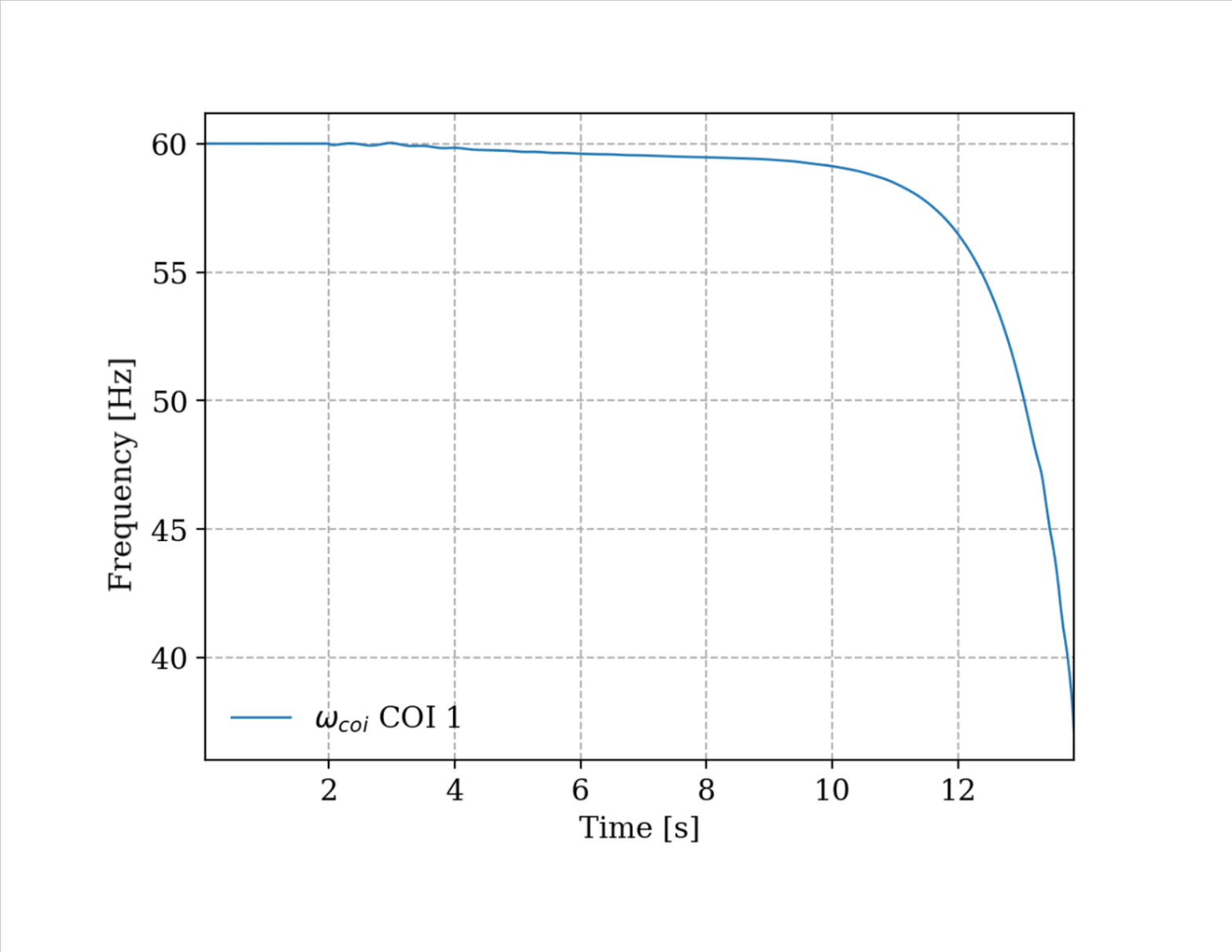}
 \vspace{-0.2cm}
\caption{The base case study of the VIS problem without $1\%$ load shedding.}
\label{fig:Fig2}
 \vspace{-0.2cm}
\end{figure}

\vspace{-0.4cm}
\begin{figure}[ht!]
\centering
\includegraphics[width=0.5\textwidth,trim=30px 70px 70px 70px,clip=true]{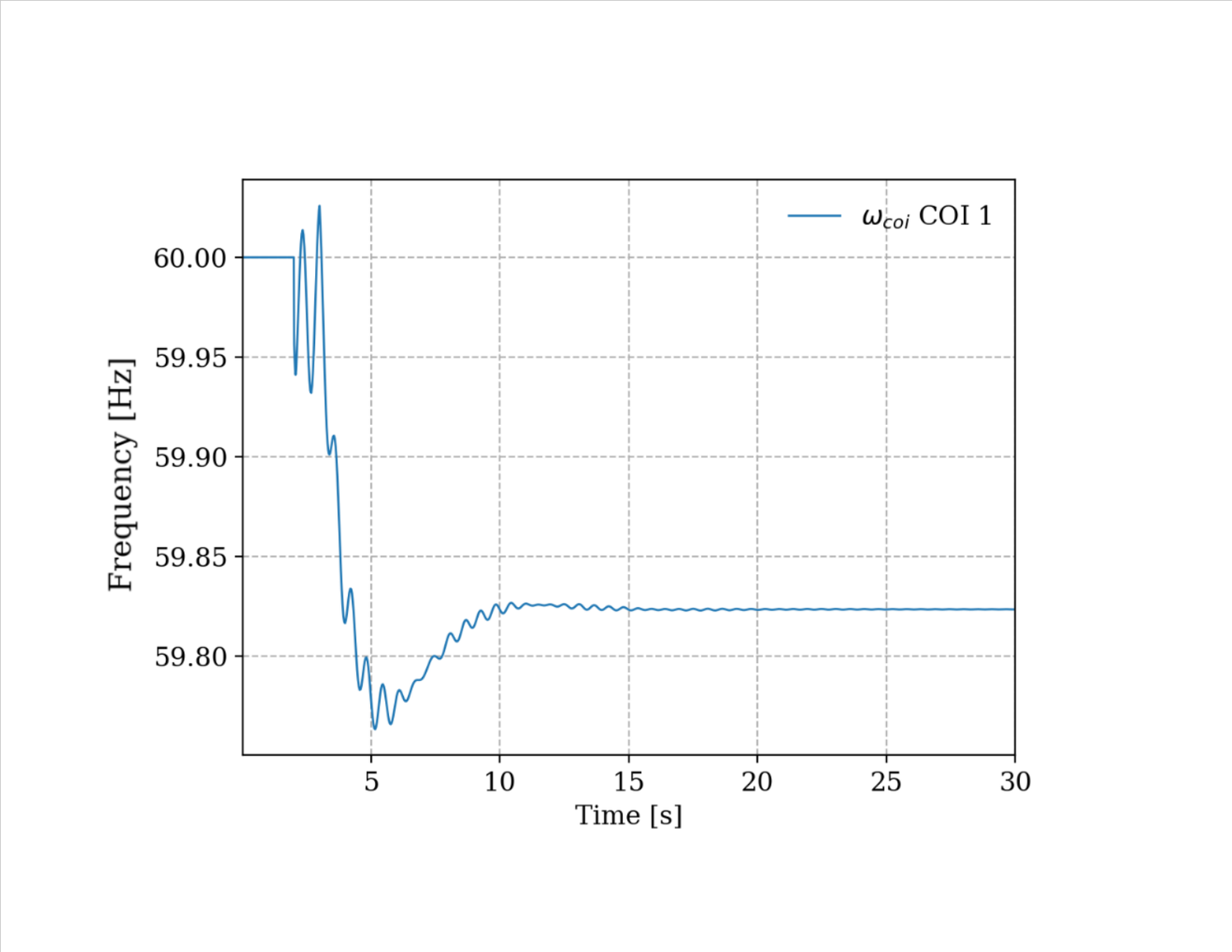}
 \vspace{-0.6cm}
\caption{The COI analysis of the proposed transient-stability-aware VIS problem without $1\%$ load shedding.}
\label{fig:Fig3}
\end{figure}
\begin{figure}[ht!]
\centering
\includegraphics[width=0.45\textwidth]{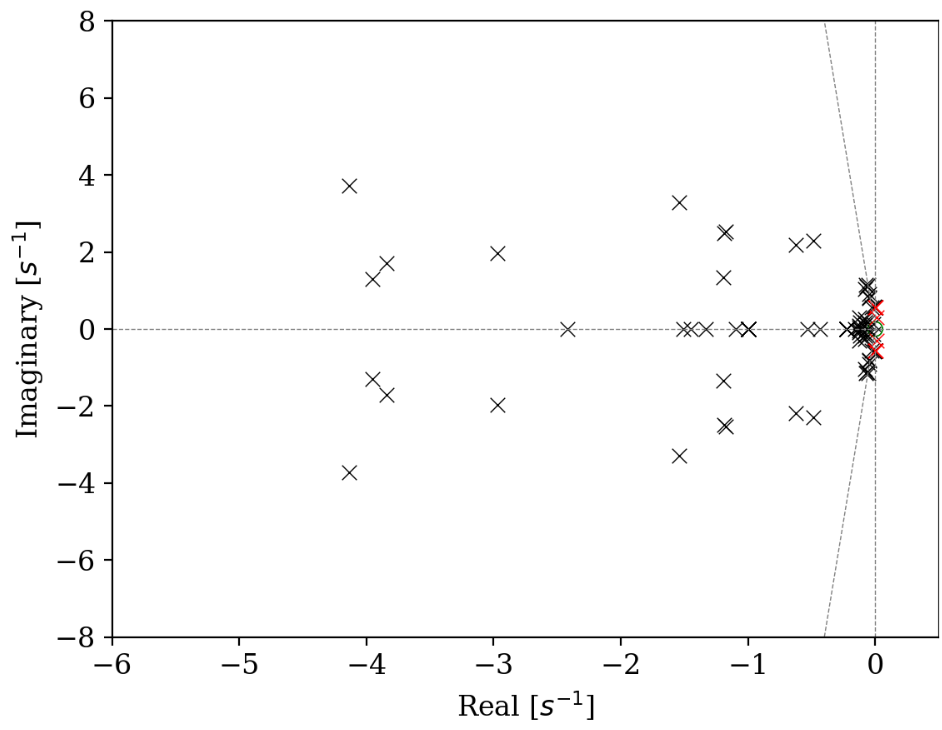}
\vspace{-0.3cm}
\caption{The eigenvalue analysis with respect to robust optimization.}
\label{fig:Fig4}
\end{figure}

Fig.~\ref {fig:Fig2} depicts the frequency disturbance during the crucial first seconds of the contingency (from $2$-$8$ s), emphasizing the vulnerabilities observed from the conventional VIS perspective. In addition, Fig.~\ref{fig:Fig2} is considered the base case of the VIS problem for the test system studied. The COI criterion clearly illustrates the limitations of the conventional VIS approach when addressing transient stability contingencies. According to Fig.~\ref{fig:Fig2}, when a fault occurs at $2$ s and a generator is lost at $3$ s, the IBR-rich grid does not return to a stable state. This failure can be attributed to the inability to accurately estimate the impact of the three-phase short-circuit fault that occurred at $2$ s. In contrast, Fig.~\ref{fig:Fig3} clearly illustrates the effectiveness of the proposed strategy, which is based on the early prediction capabilities of the DL algorithm. Once the impact is estimated using the DL strategy, a redispatch signal is sent to the committed units to prepare them for the frequency disturbance characterized by the first swing equation occurring after $2$ s. As a result, the IBR-rich grid experiences a frequency incident based on the DL algorithm's prediction. This leads to the grid returning to a stable state, including its COI, within $12$ s. Fig.~\ref {fig:Fig4} illustrates the compatibility of the optimal solution obtained from the transient-stability-aware scheme and its impact on the system's eigenvalues. In particular, all root loci for the IBR-rich grid are located in the left-hand plane of the $S$-domain, which verifies the reliability of the proposed scheme. 
\vspace{-.2cm}
\begin{table}[h!]
\caption{Estimation of the Demand Response Performance}
\vspace{-.5cm}
\label{Table3}
\begin{center}
\begin{tabular}{c|c|c|c}
\hline
\hline
\textbf{Methodology} & $\text{Cost}^{\text{total}}$ [\$] & $\text{Cost}^{\text{UFLS}}$ [\$]& $\text{Cost}^{\text{total}}+\text{Cost}^{\text{UFLS}}$ [\$]\\
\hline

\textbf{VIS} & $396.59 $ & $409.94$ & $\mathbf{806.53}$ \\
\hline
\textbf{Proposed} & $418$ & $0$ & $\mathbf{418}$\\
\hline
\hline
\end{tabular}
\end{center}
\vspace{-.2cm}
\end{table}
Table~\ref{Table3} highlights the advantages of our approach, which achieves the lowest net cost, reduces the cost associated with the IBR-rich grid during transient instability and OOS conditions, and features UFLS costs ($\text{Cost}^{\text{UFLS}}$). Importantly, our DL-based strategy incurs only a $5\%$ higher cost compared to the VIS problem, allowing the system to completely mitigate the effects of the OOS condition and regain stability after a frequency incident.

\section{Concluding Remarks}\label{sec:sec5}
This study successfully developed and validated a novel framework to manage transient stability in power systems with high IBR penetration by synergistically integrating a predictive DL model with IGDT. This approach creates a proactive, risk-averse dispatch strategy that successfully stabilizes the grid under severe contingencies where conventional VIS methods would otherwise fail. By anticipating post-fault dynamics, the framework preemptively redispatches resources to maintain the system's center of inertia within stable limits, as verified by case studies and eigenvalue analysis. The proposed method demonstrates a highly favorable tradeoff, preventing a system collapse for a marginal increase in operational cost. This work provides system operators with a practical risk-aware tool that transitions frequency control from a reactive to a predictive and stability-aware process, enabling the secure integration of IBRs at high penetration levels.

\ifCLASSOPTIONcaptionsoff
\newpage
\fi

\bibliographystyle{IEEEtran}
\bibliography{IEEEabrv,References}

\end{document}